\documentclass{article}

\usepackage{PRIMEarxiv}

\usepackage[utf8]{inputenc} 
\usepackage[T1]{fontenc}    
\usepackage{hyperref}       
\usepackage{url}            
\usepackage{booktabs}       
\usepackage{amsfonts}       
\usepackage{nicefrac}       
\usepackage{microtype}      
\usepackage{lipsum}
\usepackage{graphicx}
\usepackage{stfloats}
\usepackage{amsmath}
\usepackage{footmisc}
\usepackage{mathtools}
\usepackage{caption}
\fnbelowfloat
  
\title{A Cost Analysis of Generative Language Models and Influence Operations}

\author{
  Micah Musser \\
  Center for Security and Emerging Technology \\
  Georgetown University \\
  Washington, D.C. \\
  \texttt{Micah.Musser@georgetown.edu} \\
}

\begin{document}
\maketitle

\begin{abstract}
Despite speculation that recent large language models (LLMs) are likely to be used maliciously to improve the quality or scale of influence operations, uncertainty persists regarding the economic value that LLMs offer propagandists. This research constructs a model of costs facing propagandists for content generation at scale and analyzes (1) the potential savings that LLMs could offer propagandists, (2) the potential deterrent effect of monitoring controls on API-accessible LLMs, and (3) the optimal strategy for propagandists choosing between multiple private and/or open source LLMs when conducting influence operations. Primary results suggest that LLMs need only produce usable outputs with relatively low reliability (roughly 25\%) to offer cost savings to propagandists, that the potential reduction in content generation costs can be quite high (up to 70\% for a highly reliable model), and that monitoring capabilities have sharply limited cost imposition effects when alternative open source models are available. In addition, these results suggest that nation-states—even those conducting many large-scale influence operations per year—are unlikely to benefit economically from training custom LLMs specifically for use in influence operations.
\end{abstract}

\keywords{Influence Operations \and Language Models \and Cost Modeling}

\section{Introduction} 

For the past several years, experts have speculated that newly emerging large language models (LLMs) may be used by malicious actors to generate divisive, misleading, or false information for the purposes of social manipulation. \cite{CRESTreport, propasservice, CSETreport, workshop, kreps, middlebury, withsecure, sedova2} Organizations releasing such large language models have explicitly acknowledged this as a misuse risk, \cite{radford, deepmind} and some major players advocate for ``best practices'' on limiting access to large language models that include ``publish[ing] usage guidelines'' and ``build[ing] systems and infrastructure to enforce usage guidelines.'' \cite{cohere} However, other organizations and commentators have expressed skepticism that influence operations would benefit from using language models to produce content, potentially because they still require human curation or because the costs of generating disinformation content are already extremely low. \cite{gpt3, lazar, eleuther, snakeoil, releasestrats} Such uncertainty has resulted in calls to explicitly evaluate the costs of conventional influence operations as compared to AI-enabled ones, as illustrated in the following quote from a 2020 workshop:

\begin{quote}
  [M]odels like GPT-3 can be used to create false, misleading, or propagandistic essays, tweets, and news stories \textit{de novo} . . . [W]hile automated generation of disinformation may be feasible in principle, human labor may still be more cost-effective for such purposes. Others disagreed, and saw automated generation as much more cost-effective than training and paying humans to generate disinformation. Participants agreed that empirically investigating the economics of automated vs human generated disinformation is important. \cite{symposium}
\end{quote}

Despite this interest in the economics of influence operations, the topic remains underexplored, no doubt due in large part to the difficulty of assessing the economics of presently existing and highly secretive influence operations. To this author's knowledge, only one public attempt has been made to actually model the costs and tradeoffs facing influence operators deciding whether or not to make use of AI systems. \cite{hwang} This research primarily addresses the economics of deepfaked visual and audio content, with a focus on whether or not a technological ``arms race'' between detection systems and malicious actors is likely to happen.\footnote{For some discussion of the ``arms race'' dynamic in synthetic content, see \cite{armsrace}. Other research, such as \cite{swordsorshields}, has attempted to model the impact of AI-enabled phishing campaigns, but without examining whether the use of such AI tools is therefore cost effective for malicious actors.} 

This paper attempts to explore two related but different questions. First, what are the potential economic benefits of using language models to produce disinformation content, relative to human authorship? And second, what is the economic value of one possible policy intervention designed to reduce the risk of automated disinformation generation, namely, the use of monitoring controls on API-accessible models? To make progress on these questions, this paper attempts to model the costs of \textit{content generation} for an influence operator in various situations, including the use of a public (open source) language model, a private, API-accessible language model, or a manual campaign.\footnote{Note that I examine primarily two types of model access for LLMs: I discuss private, API-accessible models, in which a private entity owns and operates a model which users can query via an API but where the training data, code, and final model parameters are not available to actors outside of the entity; and I discuss public (or open source) models, by which I mean models where the model itself is hosted somewhere and can be downloaded and used on local computing infrastructure by any third party. Note that it is possible to further differentiate forms of structured model access, see \cite{shelvane, solaimanwired, solaiman}, and to consider other possible release decisions such as staged release or making models available only to researchers instead of all third parties. From the perspective of any given third party at a particular point in time, however, AI models can generally be classified as either inaccessible, accessible through an API, or downloadable.} I emphasize that this analysis focuses very specifically on the costs of content generation, which is only one part of the disinformation pipeline and may be—for some operations—less costly than other requirements facing operators, such as maintaining an infrastructure of inauthentic accounts or identifying the appropriate channels for distributing content. \cite{goldsteinpanel, sedova1}

The paper is organized as follows: Section \ref{sec:automation} discusses a simple base scenario: how much could the use of a language model save if the lanugage model required no human curation and could be deployed fully autonomously? Section \ref{sec:teams} then analyzes the more likely scenario of human-machine teams, where human curators review and approve model outputs instead of writing content themselves. In section \ref{sec:controls}, I then consider the cost imposition that could be generated by the use of monitoring controls on an API-accessible LLM available to would-be influence operators. Section \ref{sec:comparison} considers the value of monitoring controls under circumstances which permit operators to choose between the use of multiple models, including open source ones. While all analyses through Section \ref{sec:comparison} focus only on marginal costs, Section \ref{sec:fixedcosts} expands this to include an analysis of the fixed costs associated with different methods for accessing a language model. Section \ref{sec:robustness} further analyzes the robustness of the results from the preceding sections, and \ref{sec:discussion} concludes with a discussion of the implications of this research. 

This analysis is strictly focused on the use of \textbf{text-based} language models to generate short social media posts (which I will often refer to as ``tweets'' for the sake of focusing attention on a specific use case with relatively constant output lengths, though there is nothing platform-specific about this analysis). The model can generalize to other types of language content as well, such as news articles or blog posts. Newer text-to-image models have sparked analogous concerns about disinformation uses \cite{DRI, hwang, wapo}; this model in principle can also generalize to non-text-based forms of content generation, though the interpretation of some key parameters must change.\footnote{For instance, the cost to review an output might, when considering a text-to-image model, include touch-up work done by a human designer to finalize model outputs for posting.} I emphasize that the model and its usage here are meant to be ``first attempts'' at explicitly modeling the cost decision facing a malicious actor who is deciding whether to make use of a large language model; it has several key limitations (see Section \ref{sec:discussion}), but nonetheless may hopefully serve to inspire further refinements. In addition, further work may focus on explicitly analyzing the economcis of producing image- or video-based content using generative AI systems, as well as the use of audio-based AI systems for impersonating target individuals, all of which have recently seen a sharp rise in prevalence. \cite{mcafee, voicescams, videodeepfakes}

\section{Fully Automated Content Generation}
\label{sec:automation}

For an ``ordinary'' campaign, where all content is manually authored by humans, let $L$ represent the labor productivity of human authors (measured as outputs per hour) and $w$ represent the hourly wage of human authors. For simplicity, I treat both of these variables as constant over the full course of an arbitrarily long influence operation, which implies that the marginal cost of an additional output is constant and equal to $w/L$.\footnote{One objection to this framing is that in real influence operations, the cost of a marginal output declines over time due to the widespread use of time-saving tactics such as ``copypasta,'' as reported for instance in \cite{siochina}. There are two reasons to think that this objection need not require a non-linear estimation of manual costs. First, $w/L$ can be thought of as an amortized cost per output, and not as an extrapolated cost per output based on assumptions about how long it would take to write posts \textit{de novo}. The method for estimating these parameters, which relies on historical reporting about monthly wages as well as expected output per shift for operators, aligns more readily with this interpretation of $w/L$ as an amortized rate and not an extrapolated one. Second, while human propagandists do frequently resort to copypasta and other tactics, language models are less liable to do so. Insofar as copypasta remains a key way of identifying coordinated inauthentic activity, the use of a language model would therefore represent a quality improvement over human authorship not captured by straight comparisons of the cost per output of either a human or a language model. I am inclined to think that, due to reason one, the model presented here accurately captures the \textbf{cost} differential between a manual operation and an AI-augmented one, but also that—due to reason two—it is liable to understate the \textbf{quality} improvement that arises from switching to the use of a language model once doing so becomes cost-effective. \label{fnquality}} The only real difficulty when framing a manual campaign in these terms is to esimate these two values.

Estimates of either the wages paid to disinformation authors or the productivity of such works are hard to find, though some scattered pieces of information do exist, primarily in the context of Russian influence operations. (Because wages and labor productivity likely vary widely across campaigns conducted in different regions of the world, the specific estimates produced by this model can be thought of as reflective of the value of LLMs specifically for Russian influence operations, though see footnote \ref{50cent}.) In 2018, \textit{BuzzFeed News} reported that the Internet Research Agency (IRA) had posted job ads in 2014 and 2015 for ``social media specialists'' and ``content managers'' paying roughly \$2.86–9.53 per hour.\footnote{The specific figure from \cite{buzzfeed} is for 40,000 rubles per month for two different jobs posted sometime in either 2014 or 2015. The lower bound of this estimate comes from converting rubles to dollars at the lowest conversion rate within those years, converting to 2022 USD, and then assuming 240 hours of work per month (10 hours of work, 6 days a week, for 4 weeks). The upper bound comes from converting rubles to dollars at the highest conversion rate within those years, converting to 2022 USD, and then assuming 160 hours of work per month (8 hours of work, 5 days a week, for 4 weeks). The reason for the wide spread is primarily that the value of the ruble fell dramatically in late 2014.} \cite{buzzfeed} Reporting from the indepenent St. Petersburg-based publication \textit{Fontanka} in 2022 surfaced more information: a reporter who successfully interviewed for a job with ``Cyber Front Z'' (which appears to be linked in some way to the IRA \cite{vice}) was offered a job that would pay \$1.41–2.78 per hour.\footnote{The level of variation is again due to the use of 160 hours of work per month as a high-end estimate and 240 hours of work per month as a low-end estimate, as well as the fact that the value of the ruble fluctutated significantly in the 10 days between Cyber Front Z's hiring call and the publication of \textit{Fontanka}'s report.} \cite{fontanka} In addition, another (older) article from 2014 in \textit{BuzzFeed News} implies an average hourly wage for IRA employees in the range \$3.62–5.44.\footnote{The \textit{BuzzFeed} article places the total estimated budget of the IRA in 2014 at \$10 million, with half ``earmarked to be paid in cash'' (likely for employee salaries, of which the organization had 600 at the time). If we assume that these employees worked between 160 and 240 hours per month, and that all cash-earmarked funds were paid out as salaries, then the average hourly wage for IRA employees in 2014 would have fallen in the range \$3.62-5.44 after adjusting for inflation. This may slightly overstate the figure for employees who were tasked with content generation, who may have earned lower wages overall than other types of employees.} \cite{seddon} Though this source does not contain direct information about salaries, the fact that it falls within the overall range suggested by the other two sources is encouraging. 

Some of these sources also contain information about the expected output of content generators working for the IRA. The \textit{Fontanka} report noted that employees at Cyber Front Z were expected to write 200 comments on social media posts per shift, or somewhere in the range of 20–25 comments per hour, depending on the length of a shift. But the 2014 \textit{BuzzFeed} article about older IRA campaigns suggests thats operators managing Twitter accounts were only expected to tweet 5–6.25 times per hour.   

In the following models, I use Monte Carlo sampling to estimate both $w$ and $L$, treating both as random uniform distributions over the full range given by the above estimates. The smallest possible cost of a marginal output $w/L$ given these parameter ranges is therefore \$0.06, the largest possible cost is \$1.91, and the expectation of the marginal cost is \$0.44.\footnote{Conveniently, although these numbers were taken from a variety of sources related to Russian propaganda efforts, an expectation of \$0.44/post happens to align nicely with the notion of the ``50 cent army,'' the traditional term used for Chinese propagandists who were assumed to be paid roughly \$0.50 for each post they wrote. Although \cite{king} questions this estimate and suggests that most Chinese propagandists are salaried bureaucrats, the authors do not provide an alternate way of estimating the effective cost to the government of each post produced by these bureaucrats. \label{50cent}}

In the alternate case where an influence operation employs a language model to generate content fully autonomously, the only marginal costs associated with content generation are those required to run inference on a model. For its largest, most powerful language model,\footnote{Specifically, GPT-4 with an 8K context window.} OpenAI curently charges \$0.00006 per token, while Cohere charges an even lower \$0.000015 per token for generation tasks. \cite{cohereapi, openaiapi} Since I am generally considering tweets or comments on social media as the standard type of content in this threat model, I estimate the average token length of outputs at around 40 tokens, in which case the marginal cost for an additional output from these models would fall in the range \$0.0006–0.0024. If, alternatively, a threat actor uses an open source model which requires them to set up and maintain their own compute infrastructure, these costs may be higher, but it seems reasonable to estimate that, for any reasonably large operation, an operator could keep inference costs within an order of magnitude of the costs offered by major companies.\footnote{While major companies like OpenAI and Cohere benefit from very large economies of scale, they also deploy much larger models than a propagandist would be likely to run on local equipment, see Section \ref{sec:fixedcosts}.} The estimated values for the marginal inference cost $IC$ of an additional AI output therefore fall roughly in the range of \$0.0006–0.024.\footnote{It is worth noting that these infrastructure costs could vary substantially depending on factors like the size and capability of the model used by a propagandist. For instance, token generation with GPT-3.5 is already 30 times less expensive than with GPT-4, and running a small model on a single local GPU may be much cheaper still. In this work, I do not attempt to describe how model performance relates to per-token infrastructure costs in a way that would allow propagandists to identify the optimal choice of model for a given operation; instead, I simply treat the per-token infrastructure cost of \textbf{all} models as some constant but unknown value drawn from the above range. This is a significant oversimplification. However, final estimates for most values of interest using this model do not depend heavily on values of $IC$ (see Section\ref{sec:robustness}), and instead suggest that content generation costs with human-machine teams remain dominated by labor costs, not the infrastructure costs of running LLMs. Because relatively little depends on the precise estimation of $IC$, it seems adequate to draw a single value from a wide range of possibilities and treat it as describing infrastructure costs for all models.}

Given these estimates, a threat model of \textbf{pure} automation will always have lower marginal costs than that of a manual influence operation. This is not surprising. In addition, if an operator must expend fixed costs $FC$ to acquire a working model (whether that means training it from scratch, stealing it, fine-tuning it for an operation, or even just familiarizing one's staff with the model's capabilities), then the model pays for itself after $\frac{FC}{w/L - IC}$ outputs. With expected values $E(\frac{w}{L}) \approx 0.44$ and $E(IC) \approx 0.01$, then the use of an AI model would pay for itself after a campaign of size $2.33 FC$.

\section{Human-Machine Teams with Unrestricted AI Access}
\label{sec:teams}

With current models, it is unlikely that in \textbf{most} cases, an operator would choose to run a purely automated campaign.\footnote{However, note that if the goal of a campaign is distraction, such that the quality of individual posts does not matter to the operator, pure automation may be a perfectly workable strategy for existing language models. See \cite{king} for an analysis along these lines in the context of Chinese influence operations.} For most campaigns, especially those where the consistency and quality of posts matters heavily to the campaign's overall success, a human-machine team is a more realistic scenario. For the purposes of this paper, I imagine that a human-machine team operates in the following way: a language model is tasked with outputting content which is subsequently reviewed by a human prior to posting. The human must approve an output (perhaps with some light editing) before it is posted online. 

To incorporate an operation along these lines into this model, I introduce two additional parameters. First, let $\alpha$ represent some constant proportion indicating how much faster a human can review outputs rather than writing them from scratch, such that $\alpha L$ represents the total number of posts a human can generate and review in an hour.\footnote{``Reviewing outputs'' here includes the time necessary to prompt an LLM to generate candidate outputs for review.} And second, let $p$ represent the proportion of outputs from a language model that are usable for an operator's campaign (or that will be usable after a light edit during the review process; $p$ then falls in the range [0,1]).\footnote{Note that $\alpha$ and $p$ will be inversely related if the actual underlying capability of a given model remains constant: reviewers can choose to spend more time editing potential AI outputs or engaging in careful prompt engineering, thereby increasing the proportion of outputs that are considered ``usable'' at the cost of reducing $\alpha$, or they can simply make binary yes-no rulings on potential outputs, which increases $\alpha$ at the cost of reducing $p$. For any given combination of model and campaign, there is likely some optimal level of investment that reviewers should make in each output, but this would be hard to predict \textit{a priori}. This model attempts to handle this ambiguity by sampling from a relatively wide range of values for $\alpha$ while treating $p$ in most places as an entirely free-floating variable. But it is important to emphasize that readers trying to imagine plausible values of $p$ for existing models should \textbf{not} interpret this parameter as corresponding only to the proportion of outputs from language models that are perfectly suited for use in an influence operation with no editing or prompt engineering whatsoever.} Then the cost of producing a marginal output using a human-machine team can as be modeled as a constant, with this strategy being cheaper than paying a human to write a marginal output whenever the inequality 

\begin{equation}
  \left( \frac{w}{\alpha L} + IC \right) \frac{1}{p} < \frac{w}{L}
  \label{eq:inequality}
\end{equation}

obtains. Note that, because the inference costs of running a model are generally dwarfed by labor costs, this inequality loosely approximates to the inequality $\alpha > 1/p$, which states that whenever the speedup in a human's ability to produce and review AI generations (compared to writing them manually) is greater than the number of AI generations necessary to find a ``usable'' output, we expect the marginal cost of an output from a human-machine team to be cheaper than the marginal cost of a human writing an additional output. 

Choosing an appropriate value for $\alpha$ is one of the more difficult tasks associated with parameter estimation in this model. Although some economic studies on the labor impacts of large language models have begun to emerge, \cite{brynjolfsson, codemodels, gptsasgpts, github, korinek, mit, googleproductivity} they are mostly of limited usefulness. \cite{codemodels} and \cite{korinek} speculate that large language models will enable efficency gains for human workers but do not measure such gains, while \cite{gptsasgpts} analyzes worker exposure to large language models but not efficiency impacts of the models. \cite{brynjolfsson} estimates a 14\% efficiency improvement among call center workers using large language models, while \cite{mit} estimates a 37\% reduction in time spent on various tasks among college-educated professionals. However, these papers do not provide information about the rate at which workers reject LLM suggestions, which is necessary to calculate the costs of generating outputs but is not necessary to analyze impacts on worker productivity.\footnote{In this model, the efficiency speedup associated with the use of an LLM is disaggregated into an increase in the rate at which humans can generate and review outputs, compared to manually writing them ($\alpha$), offset by the percentage of outputs that are actually usable ($p$). This disaggregation is necessary when evaluating operator costs, because inefficiencies caused by lowering $p$ generate increased inference costs, while inefficiences caused by lowering $\alpha$ do not. In addition, the disaggregation separates efficiency gains into a parameter specific to a given human-task pair (which I estimate using Monte Carlo sampling), and a parameter specific to a given model-task pair (which I primarily treat as a free-floating variable).} \cite{github} analyzes efficiency gains for a specific code completion task and provides an absolute minimum estimate of $\alpha \approx 2.27$, while \cite{googleproductivity} estimates $\alpha$ at 4.26.\footnote{\cite{github} measures only the observed efficiency improvement on a given coding task and does not estimate $p$, similarly to \cite{brynjolfsson}. \cite{googleproductivity}, however, estimates a total efficiency gain of 6\% from the use of AI \textbf{and} finds that 25\% of AI suggestions were accepted by coders, implying a value of $\alpha = 4.26$. \cite{meta} does not estimate the total efficiency gain of using AI to assist with coding tasks but does observe that roughly 22\% of AI-generated suggestions were accepted by coders in a large-scale deployment. Because this is consistent with the acceptance rate observed by \cite{googleproductivity}, it is likely more accurate to say that \cite{github} implies a value of $\alpha$ between 9 and 10.5 when assuming a corresponding value for $p$ of roughly 0.22–0.25. \label{efficiency}}

\begin{figure}[h]
  \centering
  \makebox[\textwidth][c]{\includegraphics[width=1.1\textwidth]{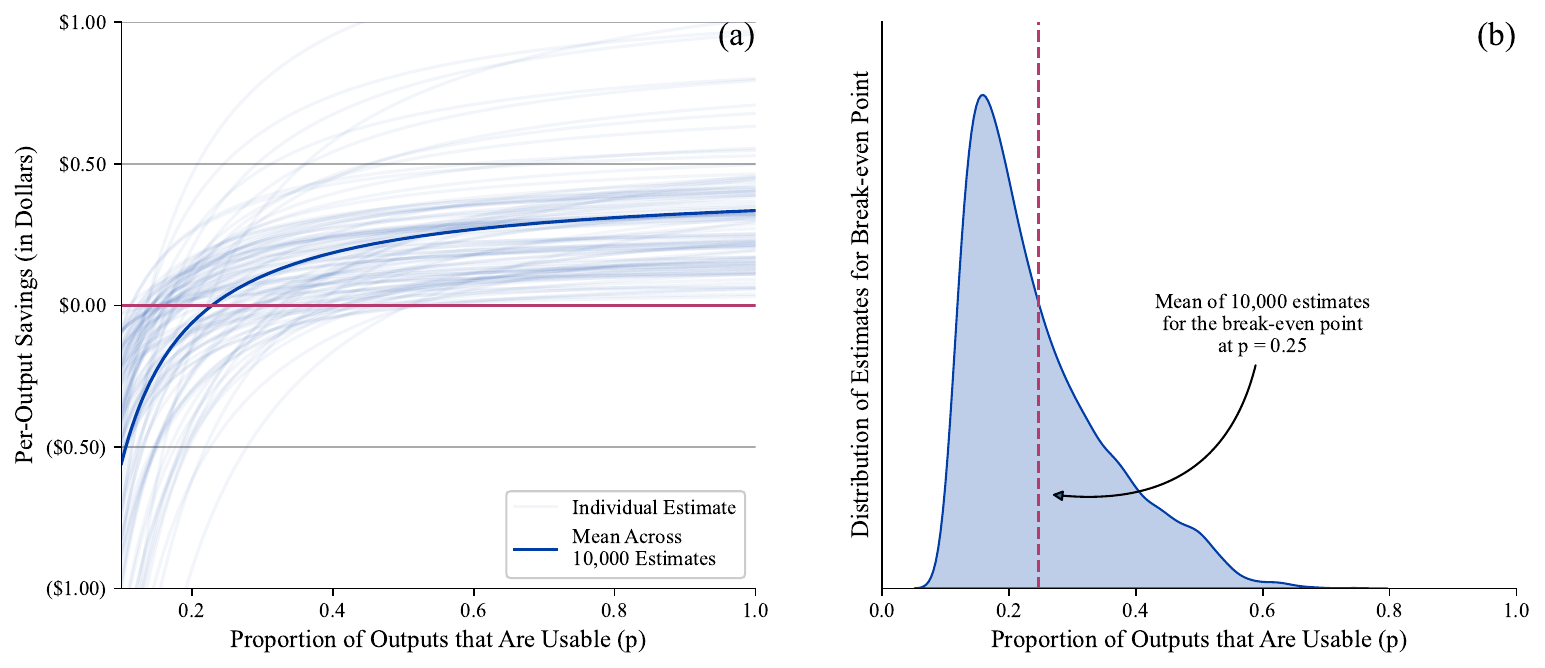}}
  \caption{Predicted Per-Output Cost Savings as a Function of $p$}
  \label{fig:costsavings}
\end{figure}

For the purposes of this paper, I randomly sample values for $\alpha$ uniformly from the range $[2, 10]$, indicating a wide range of uncertainty about how much faster it would be for an operator to generate and review outputs instead of manually writing them.\footnote{The ends of this range do, however, correspond roughly to the minimum and maximum estimates of $\alpha$ provided or implied by productivity improvement observed in specific code completion tasks with the use of AI models as discussed above.} Figure \ref{fig:costsavings}(a) samples 10,000 possible parameter estimates for each parameter except $p$ and plots the cost savings of a marginal output that could be gained from switching from manual authorship to a human-machine team as a function of $p$. In Figure \ref{fig:costsavings}(a), each thin blue line represents this cost savings as a function of $p$ for one particular choice of parameter estimates, and the thick blue line represents, for each value of $p$, the mean cost savings for a marginal output. For each of these individual sets of parameter estimates, the value of $p$ at which it becomes cost-effective to use a human-machine team instead of manual authorship is different; Figure \ref{fig:costsavings}(b) shows the distribution of break-even performance thresholds across all 10,000 parameter estimates. On average, $p = 0.25$ is the point at which an LLM is expected to become cost-effective, relative to fully manual authorship.

\begin{figure}[h]
  \centering
  \includegraphics[width=\textwidth]{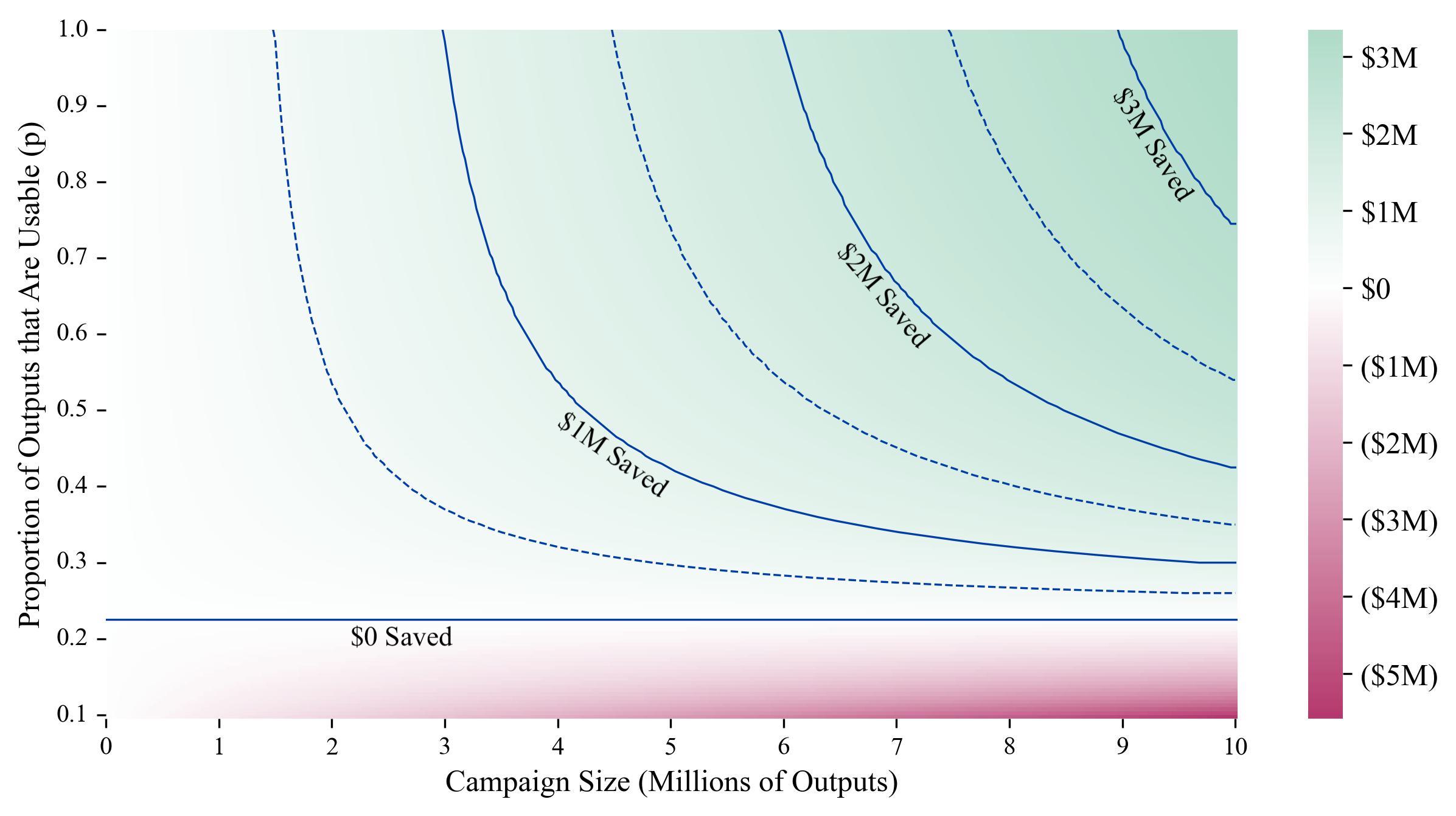}
  \caption{Cumulative Savings as a Function of Campaign Length and $p$}
  \label{fig:cumulative}
\end{figure}

Over a sufficiently long campaign, small per-output savings can add up to relatively significant amounts. Figure \ref{fig:cumulative} shows cumulative savings as a function of both $p$ and campaign length, up to 10 million tweets. (Solid lines designate savings of one million dollar increments, with dashed lines designating increments of \$500,000; savings are calculated as the mean savings for each combination of $p$ and campaign size over 10,000 parameter samples.) It is worth emphasizing that for multiple nation-states, the posting of several million tweets to Twitter is an entirely realistic goal in the medium-term. Based solely on publicly released datasets of coordinated inauthentic activity on Twitter, actors affiliated with the following countries all appear to have posted multiple millions of inauthentic tweets prior to December 2021: Serbia (17M), Saudi Arabia (17M), Turkey (15M), Egypt (7M), Iran (5.5M), Russia (5M), the United Arab Emirates (4.9M), China (3.9M), Venezuela (3.8M), and Cuba (2M).\footnote{See \cite{twitterops}. These figures were calculated based on the file sizes of ``Tweet Information'' files for each campaign, which contain metadata about tweets. As a baseline, the 353MB file corresponding to the Russian campaign released in June of 2020 contains 1.04 million tweets, suggesting that roughly 340MB of data corresponds to 1 million tweets. Note that one 4.2GB file was attributed to a joint campaign between Saudi Arabia, the United Arab Emirates, and Egypt; for simplicity, I simply divided the (imputed) number of tweets in this campaign evenly across all three countries, though it is likely given the objectives of the campaign that a disproportionately larger number of tweets came from Saudi Arabia.} These estimates are based only on infrequently released Twitter data partially covering October 2018–December 2021 (with long gaps between some releases), and is therefore likely a major undercount not only of inauthentic state-affiliated activity on Twitter specifically, but even more so of state-affiliated influence operations generally.\footnote{However, note that some of these campaigns used heavy automation; for instance, a large quantity of the posts associated with Egypt and Saudi Arabia were automated postings from the Quran. The use of LLMs to generate content compared to such heavily automated activity would likely not similarly lower costs of content generation, though it would significantly improve quality. I thank Renée DiResta for this point.}

If influence operators had unrestricted access to an LLM capable of producing usable text at least 75\% of the time, this model predicts that an operator could save upwards of \$3 million over the course of a 10-million tweet campaign, with an expected reduction in per-output content generation costs over 67\%. Moreover, based on public information about Twitter takedowns, there is a meaningful number of nation-state actors who appear likely to produce >10 million tweets (or the equivalent amount of text on other platforms) in the near- to medium-term.

\section{Monitoring Controls on AI Models}
\label{sec:controls}

Not all language models can be accessed by potential propagandists without restrictions. In the case of ChatGPT, for instance, the model itself continues to be held privately by OpenAI, with users of ChatGPT being required to make an account in order to access the model.\footnote{Note that there are many downstream applications of ChatGPT that may not require end-users to sign up for API access. However, the creator of the downstream application must themselves maintain API access to ChatGPT, and could potentially have such access revoked if their users appear to be abusing their indirect access to the original model.} Early beta users of GPT-3 were required to provide a description of their intended uses of the model prior to being granted access, but roughly eighteen months after the model's announcement, OpenAI removed the waitlist and allowed more immediate access to the model; OpenAI followed a similar trajectory with its text-to-image model DALL·E 2 after only five months.\cite{dallewaitlist, gpt3waitlist} While the model has become increasingly available to anyone to use, the fact that it remains behind a closed API makes it possible for OpenAI to monitor user interactions with ChatGPT. This monitoring is optional for a company that controls an API-accessible model, but is likely to primarily consist of using automated systems or spot checks to analyze the API requests made by users in order to assess whether a user is deliberately generating a large quantity of harmful content, where the monitoring is performed either in-house or by contractors. Users who are deemed to be deliberately generating such content could have their access to the model revoked via a revocation of API access tokens, IP address blocking, or some other measure.\footnote{Monitoring controls may also entail behavioral analysis, though this is likely to play a smaller role as compared to the detection of malicious behavior on social media platforms themselves due to the absence of user networks. In addition, companies with API-accessible models can adopt other restrictions, such as limiting the volume of allowed generations in a  given time period or blocking access from users in certain countries. I do not directly analyze the cost imposition of such controls here, though it would be possible to do so using this model: all that would be necessary is to estimate the penalty required to evade such restrictions (e.g. by creating a second account or using a VPN to avoid country-based controls) and the frequency with which such a penalty is imposed.

Note also that some forms of monitoring for open source models are also possible; for instance, it may be feasible to track information about who downloads various models hosted by third party entities, or to restrict the ability to download such models to trusted individuals. However, once a model has been successfully downloaded, monitoring of how it is used becomes impossible, and it becomes similarly impossible to impose penalties for misuse. As such, I do not consider these other forms of monitoring on open source models in this paper, and simply assume that malicious actors would be able to download an open source model and use it without interference if desired. \label{opensourcemonitoring}}

Can such monitoring controls impose meaningful costs on propagandists attempting to use language models to conduct large-scale influence operations? While blocking a user account or IP address imposes a penalty on a malicious actor, propagandists can generally create a new account or use a new IP address to continue accessing the same model, at which point the detection process must restart. In other words, if there is a roughly constant rate of detection per output $\lambda$, and each detection $D$ incurs a penalty $P$ and resets the clock for the next detection, then the costs imposed by monitoring controls over a given campaign length can be modeled as a random draw from a Poisson distribution of detections, multiplied by the penalty for each detection. Then the costs $C$ of a campaign of size $n$ will be equal to the minimum of either the manual cost of producing content, or the cost of using a language model to generate content plus the costs of evading detection:\footnote{Note that $n$ here refers to the number of \textit{usable} outputs that have been produced by the model, but since all outputs (usable or not) contribute to the model owner's ability to detect malicious use, $n$ must be divided by $p$ in equation \ref{eq:poisson} to account for unusable outputs that nonetheless contribute to the eventual detection of the malicous use.}

\begin{equation}
  C(n) = \min \left( \frac{nw}{L}, \quad \frac{n}{p} \left( \frac{w}{\alpha L} + IC \right) + P * D \sim \text{Pois}\left( \lambda \frac{n}{p} \right) \right)
  \label{eq:poisson}
\end{equation}

The penalty paid for a detection could conceivably be quite low, if a human must simply generate a new email account and sign up for the API again. Even so, doing so may generate friction costs as the human switches between reviewing outputs to creating a new account. Companies controlling access to an API-accessible model may also adopt relatively more stringent deterrance methods, for instance by requiring proof-of-humanness to sign up for an API. I—perhaps generously—imagine that each detection could require between 0.5 and 2 hours of a worker's time to evade before an operation can resume. This means that $P \sim U(0.5w, 2w)$, where $w$ itself is sampled from a uniform random distribution.\footnote{In dollar terms, $E[P] = \$6.84$.} 

\begin{figure}[h]
  \centering
  \includegraphics[width=0.75\textwidth]{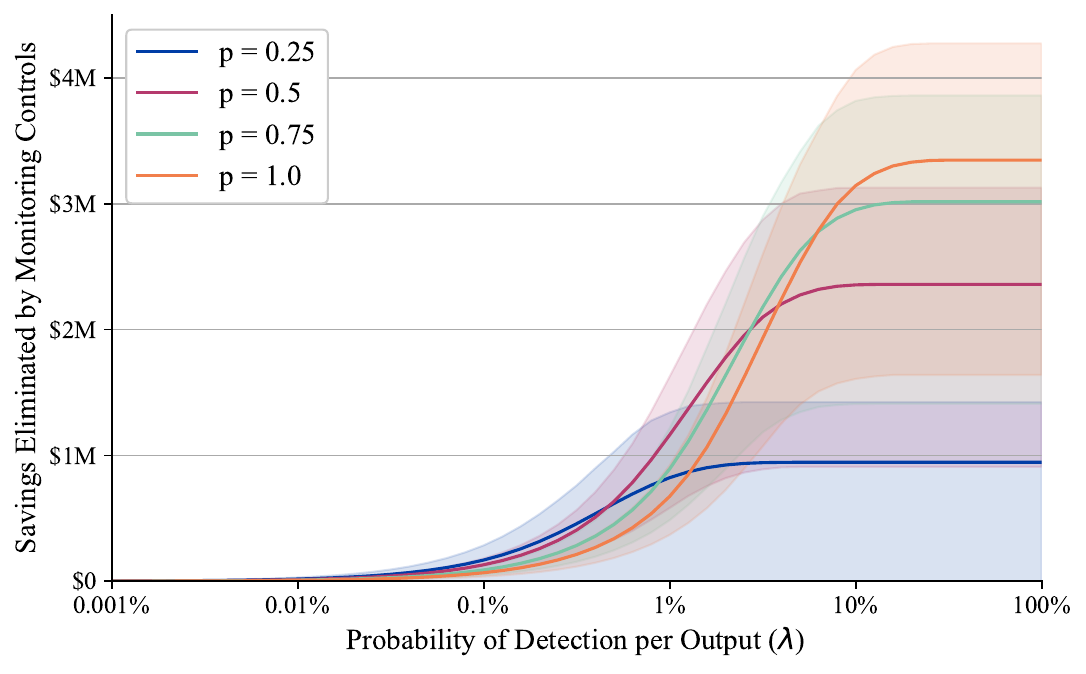}
  \caption{Penalties Imposed by as a Function of Monitoring Efficacy, for Varying Levels of $p$}
  \label{fig:penalties}
\end{figure}

Figure \ref{fig:penalties} shows, for four possible values of $p$, how improvements in detection capabilities alter the costs imposed by monitoring controls. The figure suggests that improving detection capabilities has different effects over three general phases:

\begin{enumerate}
  \item If the probability of detection per output is less than roughly 0.1\%, the monitoring controls impose minimal costs. Improvements in the ability to monitor the model do not substantially alter the cost calculus that propagandists perform.
  \item As capabilities improve from 0.1\% probability of detection per output to roughly 1\% probability of detection, costs imposed on propagandists increase by roughly similar dollar values regardless of the underlying model's capabilities. 
  \item Somewhere between a 1\% probability of detection per output and a 10\% probability of detection per output, the monitoring controls impose costs equivalent to the difference between a manual campaign and an AI-augmented one. At or above this detection rate, the propagandist prefers to use a manual campaign, and further improvements in detection capabilities impose no additional costs. The total costs imposed by monitoring controls are significantly greater for better-performing models (as the potential savings from the use of the AI model were originally much larger), but better detection capabilities are required to fully impose such costs.  
\end{enumerate}

The shaded regions in Figure \ref{fig:penalties} represent the interquartile range of possible outcomes across 10,000 parameter estimates. There is clearly enormous variation in estimates for the dollar value of costs imposed by monitoring controls, which is further analyzed in Section \ref{sec:robustness}. The general transition phases, however, are consistent across nearly all samples: a detection capability of 10\% probability of detection per output completely eliminates the incentive to use a model, while detection capabilities in the range 0.1–1\% still impose meaningful costs. 

\section{The Value of Monitoring when Public Models Are Accessible}
\label{sec:comparison}

The preceding section imagines that a propagandist must decide whether to produce content using a monitored language model or a manual process of human authorship. This might be plausible if there were a single (API-accessible) language model available to propagandists; in the real world, however, many language models have proliferated rapidly. \cite{workshop, solaiman} In this section, I instead imagine that a propagandist can choose between the use of two different language models, where model 1 is an API-accessible model with monitoring controls in place, and model 2 is an open source model instead. For now, I examine only the \textbf{variable costs} associated with either generation strategy, though in the next section I briefly discuss the fixed costs associated with downloading, fine-tuning, and running an open source model. Assuming that both the API-accessiblle and the open source model satisfy inequality \ref{eq:inequality}, the propagandist would prefer to use the private model 1 so long as the condition 

\begin{equation}
  \label{eq:public}
  \left( \frac{w}{\alpha L} + IC_1 \right) \frac{n}{p_1} + P * D \sim \text{Pois}\left( \lambda \frac{n}{p_1} \right) < \left( \frac{w}{\alpha L} + IC_2 \right) \frac{n}{p_2}
\end{equation}

is satisfied. To make this equation slightly more manageable, we can assume that the inference costs are the same regardless of model.\footnote{This decision is justified by the fact that inference costs are generally dwarfed by labor costs in this model; see Section \ref{sec:robustness}. \label{icequivalent}} From the propagandist's perspective, where $D$ is unknown at the start of the campaign, we may also substitute $D \sim \text{Pois}\left( \lambda \frac{n}{p_1} \right)$ with $E[D]$, which is just $\lambda \frac{n}{p_1}$. Then, with some rearranging, inequality \ref{eq:public} becomes:

\begin{equation}
  \label{eq:public2}
  P\lambda < \left(\frac{w}{\alpha L} + IC\right)\left(\frac{p_1 - p_2}{p_2}\right)
\end{equation}

Inequality \ref{eq:public2} states that the propagandist's expected marginal costs from relying on an API-accessible LLM are lower than those of the open source LLM only if the penalty per detection times the detection rate is lower than the marginal cost of reviewing an output, times the percentage performance improvement that the private model offers relative to the public one.\footnote{Note that if the open source model is actually better-performing than the API-accessible model, the right-hand side of inequality \ref{eq:public2} will be negative. Since the left-hand side is necessarily positive, this means that the API-accessible model is never preferred to the open source model when considering only \textbf{variable} costs. However, as Section \ref{sec:fixedcosts} briefly discusses, it is possible for a propagandist to prefer a worse-performing API-accessible model over a better open source one if the fixed costs associated with the open source model are sufficiently high.} 

Let $\hat{p}$ represent the threshold performance of an AI model at which it becomes cost-effective to use the model, relative to a manual campaign. Then there are four relevant scenarios that determine the propagandist's cost-optimal strategy: 

\begin{enumerate}
  \item If $p_1 \leq \hat{p} \land p_2 \leq \hat{p}$, the propagandist prefers to use a manual campaign regardless of any monitoring controls on the API-accessible model;
  \item If $p_2 > \hat{p} \land p_2 > p_1$, the propagandist prefers to use the better-performing open source model regardless of any monitoring controls on the API-accessible model; 
  \item If $p_1 > \hat{p} \land p_2 \leq \hat{p}$, the propagandist prefers to use the API-accessible model, but will fall back to the use of a manual campaign if monitoring controls impose sufficient costs; and
  \item If $p_2 > \hat{p} \land p_1 > p_2$, the propagandist prefers to use the API-accessible model, but will fall back to the use of the open source model if monitoring controls impose sufficient costs.
\end{enumerate}

\begin{figure}[h]
  \centering
  \includegraphics[width=0.9\textwidth]{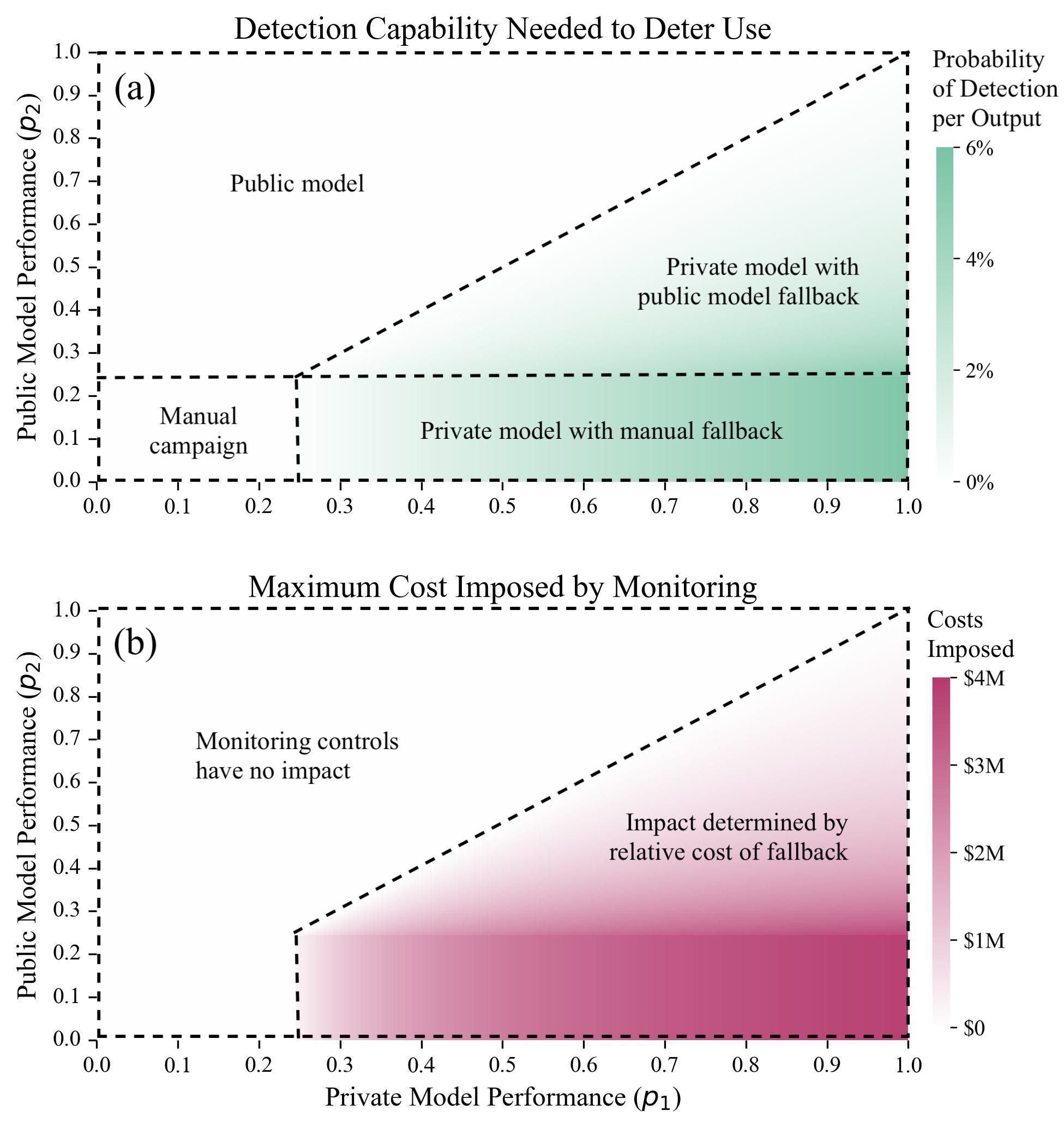}
  \caption{Optimal Strategies and Maximum Costs Imposed as a Function of $p_1$ and $p_2$}
  \label{fig:strategies}
\end{figure}

For each pair $(p_1, p_2)$ that satisfies either condition 3 or 4 above, it is possible to estimate the minimum detection capability $\hat{\lambda}$ that imposes sufficient costs to deter the propagandist from using the API-accessible model. For condition 3, this value can be estimated using equation \ref{eq:poisson}, and for condition 4, it can be estiamted using equation \ref{eq:public2}.\footnote{See Appendix \ref{sec:supplemental} for equations used to calculate $\hat{\lambda}$, including when the propagandist must pay fixed costs to access and/or fine-tune a public model.} Since further improvements in detection impose no additional costs after a propagandist has resorted to their fallback strategy, the maximum cost imposition of monitoring controls over a campaign of length $n$ can further be estimated as $P \hat{\lambda} \frac{n}{p_1}$. Figure \ref{fig:strategies} shows, for all values of $(p_1, p_2)$, the optimal strategy pursued by the propagandist, the detection capability needed to make the propagandist indifferent between the use of the API-accessible model and the relevant fallback option (inset (a)), and the costs imposed by such a detection capability over a ten-million-tweet campaign (inset (b)). 

The lower right-hand rectangle of Figure \ref{fig:strategies}(b) shows similar information as Figure \ref{fig:penalties}: with ideal detection capabilities, the maximum cost imposition of monitoring controls ranges from under \$1,000,000 to roughly \$3,500,000 over the course of a ten-million-tweet campaign, depending on the value of $p_1$. However, Figure \ref{fig:strategies}(b) further shows that these cost impositions are dramatically reduced when operators can instead switch to alternative open source models, even if those models perform less well. For instance, if both models perform with $p > 0.5$, but the best available open source model consistently performs only 90\% as well as the private model, then optimal detection capabilties (roughly, a 0.2\% probability of detection per output) will impose under \$250,000 in additional costs. 

\section{Fixed Costs Associated with Running and Training Local Language Models}
\label{sec:fixedcosts}

The previous discussion focuses entirely on variable costs associated with a manual campaign, the use of an open source model, or the use of an API-accessible model. For simplicitly, I have treated the fixed costs associated with each type of campaign as negligible.\footnote{A manual campaign theoretically does require upfront costs to acquire talent and create infrastructure. In fact, these fixed costs may dwarf the actual content generation costs, because maintaining a large infrastructure of accounts is often more costly than content production. But for this cost modeling exercise, a more realistic threat scenario is that of an already-existing propaganda outlet deciding whether or not to begin producing content using language models, a scenario under which the fixed costs for the manual campaign have already been paid. There may also be fixed costs associated with the initial account creation and familiarization with the technology involved in the use of an API-accessible model, but I treat these costs as negligble. Downloading and running an open source model on local infrastructure likely carries the greatest fixed costs, but these can be incorporated into the following adjusted model; see footnote \ref{feasibilityset}, below.} However, a propagandist need not treat the performance of an open source model as fixed; instead, they can choose to expend some additional up-front resources on fine-tuning the model to improve it. Let $FR$ represent the feasibility region consisting of all points $(FC, p_2)$ for which it is feasible to reach a given performance by expending $FC$ in fixed costs.\footnote{I frame this as a question of expending fixed costs to improve a model through fine-tuning, but if there are fixed costs to running a model in the first place, this can be handled by defining $FR$ such that any point in the set $\{ FC, p_2 : FC = 0, p_2 > 0 \} \notin FR$. \label{feasibilityset}} Then the total expected costs facing the operator are given by:\footnote{Note that equation \ref{eq:totalcosts} can be expanded to include relevant comparisons of multiple API-accessible models or multiple locally-running open source models. For instance, the feasibility region of an existing, relatively small model may not include high values of $p$ which could be achieved were a propagandist to train a larger model from scratch, though such training might require much higher fixed costs. Similarly, multiple API-accessible models may exist with different detection capabilities and performances.}

\begin{equation}
  \label{eq:totalcosts}
  C(n) = \min \left(
  \begin{aligned}
    \text{Manual} &= \frac{nw}{L} \\
    \text{API-Accessible LLM} &= \frac{n}{p_{1}} \left( \frac{w}{\alpha L} + IC + P \lambda \right) \\
    \text{Open Source LLM} &= \frac{n}{p_{2}} \left( \frac{w}{\alpha L} + IC \right) + FC, \quad (p_2, FC) \in FR
  \end{aligned}
  \right)
\end{equation}

The question of defining the feasibility region, that is, of articulating what types of capabilities are possible at various levels of investment, is both task-specific and requires advanced technical knowledge that goes well beyond the scope of this paper. Nonetheless, existing knowledge can be used to make some general estimates, as in the following scenarios:

\begin{itemize}
  \item Suppose that ChatGPT-3.5 is capable of producing ``usable'' outputs for an operator at a rate of 0.85, but that ChatGPT-4 has a higher success rate of 1.0.\footnote{\cite{surveyresearch} used GPT-3 to generate articles making the same claims as a subset of known propaganda articles and compared their relative impact on readers' beliefs. Out of 18 articles generated using GPT-3, only two were deemed not relevant to the thesis statement intended by the researchers, for a success rate of 89\%. Although this is a low threshold, so too is my threshold of ``usable'' outputs. In reality, due to additional safety measures implemented by OpenAI, ChatGPT-4 is likely to actually respond to malicious requests far less often than with perfect accuracy, but because a value of 1.0 represents the maximum benefit of the model, I use it here for illustrative purposes.} Because ChatGPT-4 requires a \$20 monthly fee to access, while ChatGPT-3.5 does not, we can treat the penalty for detection from ChatGPT-4 as \$20 higher than the penalty for detection from ChatGPT-3.5 (assuming that the propagandist will be detected at least once per month, and thus effectively pays this as a one-time signup fee after each detection). Plugging these values into the line for the API-accessible LLM in equation \ref{eq:totalcosts}, using our Monte Carlo sampling for the other parameters, and rearranging, we can estimate that the propagandist will prefer to use ChatGPT-3.5 as long as $\lambda > 0.0009$ (i.e., as long as there is at least a 0.09\% probability of detection for each output). 
  \item Suppose further that OpenAI has in fact implemented monitoring controls sufficient to detect malicious action at this rate. However, the propagandist can also reach performance on par with ChatGPT-3.5 by expending only \$600 to download and fine-tune an existing open source model, i.e. the point $(\$600, 0.85) \in FR$.\footnote{In \cite{alpaca}, researchers were able to achieve GPT-3.5-level performance for less than \$600 in fine-tuning expenses. Note that the cost required to both train and fine-tune open source models in order to reach arbitrary levels of capability appears to be continuously declining; see also \cite{mosaic}. For this reason, the estimates provided here if anything underestimate the ease and economic value of quickly pretraining or fine-tuning existing open source models for use in specialized influence operations, as opposed to relying on API-accessible models.} Again plugging the relevant values into equation \ref{eq:totalcosts} and using Monte Carlo sampling for the remaining parameters, we can estimate that the propagandist will prefer to use the open source model if they anticipate using it for more than roughly 250,000 outputs. 
  \item Finally, suppose that the propagandist cannot further improve any existing open source language models beyond this threshold with additional fine-tuning. However, the propagandist can choose to train a more advanced model of their own for \$4,600,000 which could perform as well as ChatGPT-4 (but without any usage monitoring).\footnote{\cite{gpt3cost} estimates the total cost of training GPT-3 at \$4,600,000. For this scenario, I imagine that the performance of GPT-4 could be replicated with a GPT-3-sized model plus fine-tuning, but that it could not be replicated with a smaller model.} If the OpenAI models were the only ones available, training this model would be cost-effective if the propagandist planned to engage in influence operations requiring roughly 310 million outputs or more. But at that scale, using the \$600-fine-tuned open source model is more cost effective than either OpenAI model, and compared to \textbf{that} alternative, the propagandist only finds it cost-effective to train a model from scratch if they intend to conduct campaigns larger than 410 million outputs in size.\footnote{Note that, as per the final paragraph of Section \ref{sec:automation}, if the propagandist were able to produce a model that could operate \textbf{fully} autonomously, this expenditure would become cost-effective at campaigns of only 10.7 million outputs. However, this would require that there be no human involvement even for the identification of posts to comment on and the writing of prompts to generate those posts from the AI model.}
\end{itemize}

Although they require a lot of suppositions, these scenarios are useful for illustrating some general points: for very small campaigns, propagandists are likely to prefer using API-accessible models, even if those models have monitoring controls that impose significant costs. But given only moderate assumptions about the payoffs of fine-tuning small, lightweight models to perform propaganda-specific tasks, it very quickly becomes more cost-effective for operators to rely on such models. And even when those models still have relatively large limitations that necessitate continued and careful human curation of outputs, training a large language model from scratch is almost never economically worthwhile except at extremely large scales. 

\section{Sensitivity Analysis}
\label{sec:robustness}

The previous results include the following specific estimates: 

\begin{enumerate}
  \item \textbf{Threshold Performance}: A marginal output produced by a human-AI team is expected to become cheaper than a marginal output written by a human author whenever a language model is able to produce usable outputs at a rate higher than 0.25 (95\% CI: [0.12, 0.51]).
  \item \textbf{Maximum Savings}: Over the course of a 10-million-tweet campaign, with a language model that produces usable outputs at a rate of 75\%, a propagandist could expect to save \$3 million in content generation costs, on average (assuming no fixed costs to using the model and no monitoring controls in place on the model; 95\% CI: [\$430,000, \$9.4 million]).
  \item \textbf{Optimal Detection Rate (API Only)}: If an operator can access an API-accesible model that produces usable outputs at a rate of 75\%, and if their only fallback in response to costly monitoring controls is to resort to human authorship, then monitoring controls that can detect misuse with a probability of at least 4\% per output would be required to fully deter the propagandist from using the API-accessible model (95\% CI: [0.9\%, 12\%]).
  \item \textbf{Maximum Cost Imposition (Public Option)}: However, if an open source but slightly worse-performing model (say, one that produces usable outputs at a rate of 70\%) exists, then the maximum cost imposition generated by monitoring controls is \$740,000 (95\% CI: [\$44,000, \$3.0 million]).
  \item \textbf{Minimum Viable Size (Fine-tuning vs. API)}: If a propagandist can fine-tune an existing open source model for \$600 to produce usable outputs at a rate of 85\%, and if a similarly capable API-accessible model exists that with a 0.1\% probability of detection per output, then the fine-tuned model is preferred for any campaign larger than roughly 130,000 outputs (95\% CI: [38,000 outputs, 420,000 outputs]).
  \item \textbf{Minimum Viable Size (Training vs. Fine-tuning)}: However, if reaching a performance reliability of 100\% requires training an LLM from scratch at roughly the cost of GPT-3's original training run (\$4.6 million), the training from scratch is only cost effective for campaigns larger than roughly 410 million outputs (95\% CI: [82 million, 1.1 billion]).
\end{enumerate}

These estimates require a number of parameters to be manually specified, primarily models' performance rates, detection capabilities, and fixed costs (at least when these parameters themselves are not the object of analysis). However, the estimates also rely on Monte Carlo sampling for five key variables: $\alpha$, $w$, $L$, $IC$, and $P$. Table \ref{tab:parameters} provides summaries for the ranges over which values for each variable were (uniformly) sampled, and reiterates the general source(s) from which each of these ranges were extrapolated. 

While all of these sampling ranges span large regions of uncertainty, uncertainty in some parameters more strongly drive variation in estimates for the above numerical results than uncertainty in others. Figure \ref{fig:sensitivity} attempts to visualize the way that uncertainty in each parameter contributes to variation in estimates for the six numerical results listed above. Each boxplot represents 10,000 samples of a particular parameter of interest where only the parameter on the x-axis is allowed to vary; all other parameters are held constant at their midpoint value. Effectively, this samples from the marginal distribution for the estimate with respect to a single parameter of interest. The final, yellow boxplot on the right-hand side of each subplot shows the variation in the final estimate when all parameters listed along the x-axis are allowed to vary on the ranges shown in Table \ref{tab:parameters}, that is, the yellow boxplot represents the overall probability distribution for the estimate as given by this model.

Subplot (a), for instance, shows that the vast majority of uncertainty in the predicted threshold performance at which it becomes cost effective to use a language model depends on $\alpha$, and not on inference costs, wages, or labor productivity. By contrast, other parameters of interest are more heavily determined by values of $w$, $L$, and—for estimates where monitoring controls on an API-accessible model are relevant—$P$. No parameter estimate heavily depends on variation in $IC$, which further justifies treating the inference costs of different models as equivalent (see footnote \ref{icequivalent}).

Interestingly, most estimates for the potential savings that unrestricted access to a reasonably reliable LLM could generate (supblot (b)) span roughly two orders of magnitude. But the estimates for the maximum potential cost imposition of monitoring controls when a slightly-less-reliable open source model exists (subplot (d)) span a wider three orders of magnitude. There is also relatively large uncertainty regarding the detection capability needed to fully deter propagandists from using an API-accessible model (when public options do not exist, subplot (c)). There is less variation, by contrast, in the estimates regarding the scale of operation at which a fine-tuned open source model becomes cheaper than a similarly-performing API-accessible model (subplot (e)). Note that, since such a comparison assumes that the two models perform similarly, the only differences between them when minimizing equation \ref{eq:totalcosts} depend on $FC$, $\lambda$, and $P$ (which is itself partially a function of $w$)—thus, variation in $\alpha$, $L$, and $IC$ does not cause variation in the estimated scale necessary for a fine-tuned open source model to become preferred to a private, API-accessible one.

\noindent\begin{minipage}{\linewidth}

\renewcommand{\arraystretch}{1.5}

  \centering
  \captionof{table}{Sampling Ranges for Key Parameters}
  \begin{tabular}{c|c|c|c|c}
    \hline
    Parameter & Lower Bound & Upper Bound & Midpoint & Justification \\
    \hline
    $\alpha$ & 2 & 10 & 6 & Observed Values from Code Generation Tasks \\
    $w$ & \$1.41 & \$9.53 & \$5.47 & Historical IRA Job Postings and Operations \\
    $L$ & 5 & 25 & 15 & Historical IRA Job Postings and Operations \\
    $IC$ & \$0.0006 & \$0.024 & \$0.01 (approx.) & API Fees for Existing Models \\
    $P$ & 0.5$w$ & 2$w$ & 1.25$w$ & Optimistic Range of Impact \\
    \hline
  \end{tabular}
  \label{tab:parameters}

\renewcommand{\arraystretch}{1.0}

\vspace{1cm}

  \centering
  \includegraphics[width=0.95\textwidth]{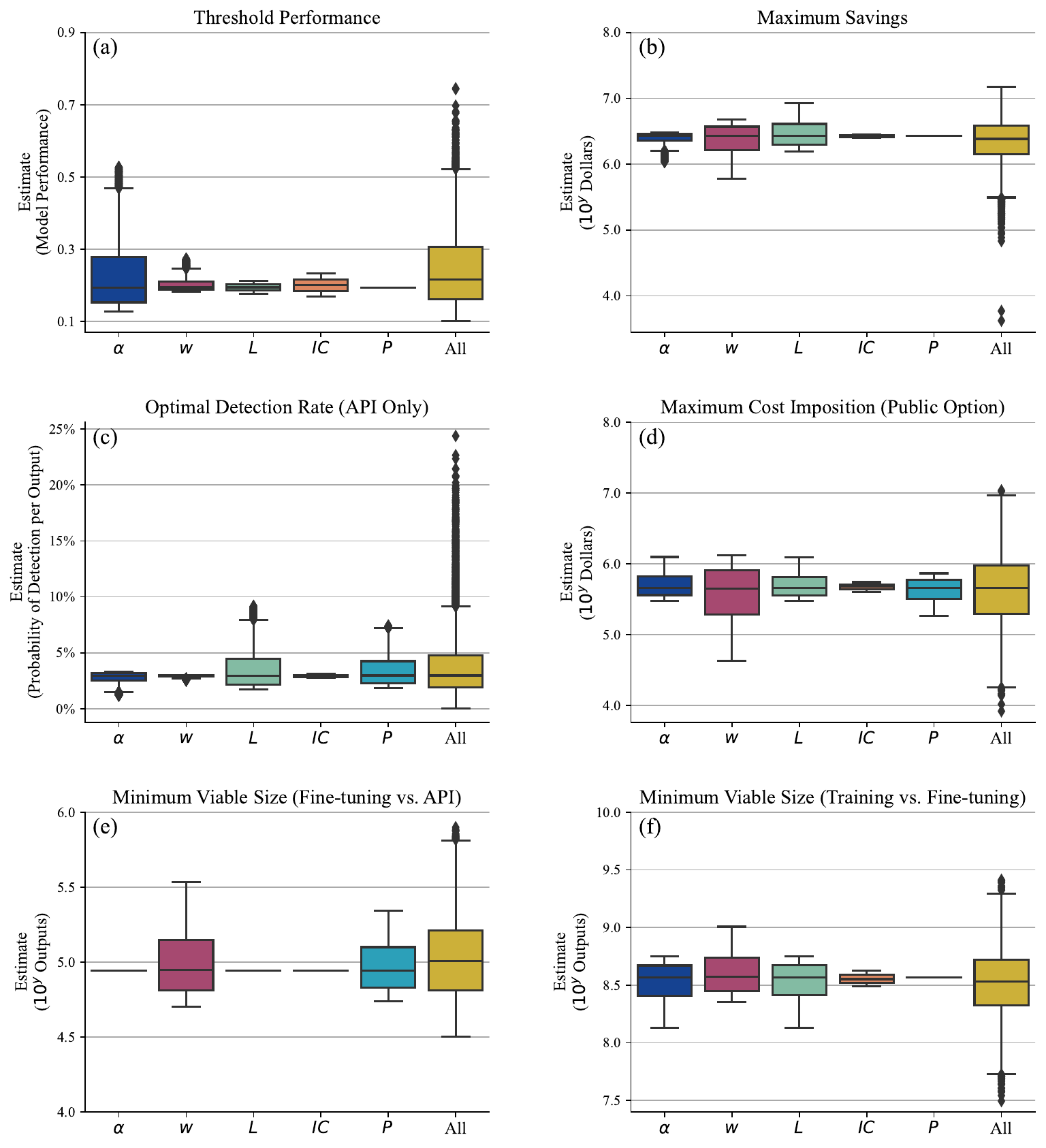}
  \captionof{figure}{Contributions from Uncertainty in Key Parameters to Variation in Overall Results}
  \label{fig:sensitivity}

\end{minipage}

\section{Discussion}
\label{sec:discussion}

The preceding analysis suggests that, under a relatively wide range of potential scenarios, the use of language models to produce misinformation content is highly cost effective, relative to the use of purely manual content generation. While it is not surprising that a fully automated campaign would be cheaper than paying humans to write content for influence operations, these models suggest that the use of even relatively unreliable models can substantially reduce propagandists' costs via human-machine teams, as long as models produce ``usable'' outputs more often than one in four times. With human-machine teams, labor costs still dominate an operation's overall content generation costs, but savings can quickly approach the millions of dollars. 

Before concluding, it is worth discussing a few general comments on the implications and limitations of this work.

\subsection{Model limitations}

This model of influence operations is limited in a number of key ways. First, and most notably, parameter estimates regarding worker productivity and wages in existing influence operations are based on a small number of investigative reports or job postings, almost exclusively in the context of Russian influence operations. These figures may or may not generalize to other propaganda operations, but an absence of public data about the organization and economic structures of propaganda campaigns makes further precision difficult. Additionally, while some research has begun to emerge regarding the impact of LLM usage on worker productivity, \cite{brynjolfsson, gptsasgpts, codemodels, mit, googleproductivity}, there is still large uncertainty regarding how effective disinformation operators will be at incorporating LLMs into their workflows. Future economic research in other domains may significantly help to narrow the uncertainties in this model.  

Relatedly, the model analyzes the economics of using LLMs for discrete tasks insofar as it uses a single value—$p$—to describe a model's capability. But $p$ is not meant to be a description of a model's abstract capabiltiies, but rather its reliability at producing usable outputs on a specific task. For instance, a model may perform reliably enough to save money on the task of tweet generation, but struggle more with longer-form content, making it cost ineffective for use on the task of fake news article generation. To \textbf{some} extent, it may be reasonable to interpret a finding that ``use of such and such a model becomes cost effective given certain assumptions at $x$ outputs'' as meaning that the model becomes cost effective when used across multiple tasks to produce any type of written content as long as $x$ tweets. But such an inference relies on the assumption that performance across tasks is relatively consistent, which may or may not be true. In other words, while the models used here can give a rough sense of the scale at which certain content production strategies become cost saving, they do not fully capture the multiplicity of content generation tasks for which actual propagandists would be likely to use LLMs. 

This model is also strictly focused on the cost savings associated with producing content at scale, and not with quality improvements that LLMs could offer propagandists. But major quality improvements may be possible, providing an additional incentive to makes use of LLMs in propaganda campaigns. The use of copypasta, stilted language from non-native speakers, or transliterations of idioms that do not make sense in a propagandist's target language have often provided important clues regarding inauthentic behavior. \cite{siochina, graphika, bjp} These errors in human cultural awareness and translation ability are more prevalent in influence operations conducted by some countries than others, and countries whose propagandists frequently make such errors may be forced to prioritize volume in output instead of taking time to carefully craft believable—and more persuasive—personas for their fake accounts. \cite{chinatwitter, dragonbridge} Language models, by contrast, are unlikely to make such easily-noticeable mistakes, though they may still struggle to effectively mimic discursive norms common in fine-grained target populations. \cite{gpt3, CSETreport, sedova2} Other research has noted that content produced from GPT-3 can change readers' opinions on sensitive political issues, and can do so even better than existing examples of propaganda news articles with only light copyediting. \cite{surveyresearch} These improvements in quality may permit propagandists to meaningfully alter the strategies that they employ in influence operations. \cite{workshop}

\subsection{Factors other than cost may disincentivize the use of LLMs in influence operations}
\label{sec:bureaucracy}

Even if it is strictly cost-effective for malicious actors to use LLMs to produce disinformation content, organizational, bureaucratic, or cultural barriers may cause propaganda outlets to nonetheless avoid doing so. Propaganda outlets can take a number of forms. \cite{fisher} describes the proliferation of private ``disinformation-for-hire'' firms that are contracted to generate and promote disinformation, whereas \cite{king} argues that large quantities of Chinese-origin propaganda on social media is produced by a diffuse set of government bureaucrats who are paid per output to produce propaganda content but without any centralized direction or oversight. Propaganda outlets organized around the first model will likely have much stronger incentives to adopt cost-saving technologies than those operating on the second model. In the extreme, we can even speculate that bureaucratic structures which reward departments on the basis of personnel size may actively disincentivize the adoption of LLMs for propaganda purposes.\footnote{It is not clear whether any major propaganda outlets face this set of incentives. However, one common issue facing propagandists is that it is remarkably difficult to evaluate the impact of disinformation on actual political attitudes and behaviors, with some research indicating that the concrete effects of exposure to influence operations is relatively small. \cite{nature} The difficulty of evaluating the political import of specific operations, combined with the insulation from cost-cutting pressures that exists for in-house propaganda authors (contrasted with specialized firms), could largely nullify the incentives to adopt LLMs for content generation.} \cite{oii} notes that many countries continue to use decentralized organizational models to produce content for influence operations, but that the current trend is increasingly towards centralized, third-party firms, which are much more likely to have the coordination and desire necessary to adopt new cost-saving technologies. According to \cite{oii}, between January 2019 and November 2020, public government contracts suggest that states have paid at least \$60 million (and likely much more) to private forms to conduct influence operations. 

More generally, it is unclear to what extent propagandists are optimizers or satisficers. The development of ``deepfake'' technology over the 2010s led some analysts to speculate that Russia would unleash a ``wave'' of deepfaked disinformation against the West. \cite{wave} However, despite some high-profile examples \cite{zelenskyy}, deepfaked images and videos remained an apparently minor component of Russian influence operations for relatively long after the technology to produce them existed. Only in recent months, with the rise of text-to-image generative AI systems, have AI-generated images become more commonly observed as a tool for disinformation (though not necessarily as a tool of Russian disinformation specifically). \cite{trump, putin} This potentially suggests that technical barriers to adoption can meaningfully deter propgandists from using new technologies, and that improvements in user-friendliness affect propagandists' decision-making more than improvements in underlying capabilities. 

Similarly, although the use of LLMs to produce disinformation was largely speculative until recently, recent months have seen networks of Twitter accounts posting tweets containing ChatGPT's default refusal to complete a user request response, likely suggesting attempts to use the model to generate content to post on social media. \cite{conspirator} While the models presented here suggest that for even relatively small campaigns, a propagandist's cost-optimal solution is to fine-tune an open source model, propagandists may instead prefer to rely on solutions with lower adoption costs even when doing so is economically irrational. 

\subsection{Nation-states do not have strong incentives to secretly train LLMs for influence operations}

The maximum length of a campaign evaluated in the preceding sections was 10 million tweets. This volume of coordinated inauthentic activity on Twitter between October 2018 and December 2021 was exceeded by only three countries (see Section \ref{sec:teams}). However, this is true only when considering (1) publicly attributed activity that was (2) posted to and removed by Twitter (3) over a three-year period with significant gaps in reporting. It seems reasonably likely that for a small but significant number of nation-state actors, the amount of content generated for use in influence operations over the near- to medium-term could substantially exceed the equivalent of 10 million tweets. In fact, \cite{king} estimates that the Chinese government fabricates and posts ``about 448 million social media comments a year.'' 

However, even at this scale, the value of training an LLM from scratch to produce disinformation content—as opposed to simply fine-tuning an existing open source model—is dubious, even if the fine-tuned model is not as capable (see Section \ref{sec:fixedcosts}). If the best attainable performance of any open source model, even after careful fine-tuning, was still very low for a given task, it \textbf{might} become economically viable to train an LLM from scratch. But the propagandist must not only believe this to be true of existing open source models, but also of any future models that may be released between the time the operator begins training their model and the time they generate enough posts to fully recoup their expenses. Given the rapid rate of both public model release and the tendency of access to privately-held models to become easier and less restricted over time, this is unlikely to be a reasonable bet.\footnote{Note also that, even if the Chinese government produces roughly 448 million inauthentic social media posts per year, this volume of content is likely not produced by a single propaganda agency that could pay the initial fixed costs of model training and then recoup their expenses over time; rather, it is produced (at least in part) by a diffuse set of bureaucrats, no one member of which stands to gain from paying large upfront costs for the sake of increasing their individual efficiency at generating misinformation. \cite{king}} 

It is possible that the use of LLMs may themselves enable much larger campaigns, such that although training a model from scratch would be a poor economic decision under \textbf{current} scales of operation, doing so would enable much larger scales of operation that \textbf{do} justify such an investment. But there are checks on the scale at which propagandists can operate that go beyond the costs of content generation, including the difficulty of maintaining large networks of inauthentic accounts without being detected by platforms. \cite{goldsteinpanel, sedova1} It seems reasonably likely, then, that even nation-states may find it difficult to justify secretive large-scale training runs of LLMs intended primarily for use in influence operations. 

\subsection{The comparative value of technical mitigations against LLM misuse}

There are three broad technical intervensions which LLM developers can pursue to reduce the likelihood of their models being abused: they can train or fine-tune the model itself in ways that reduce its propensity to comply with malicious user requests (thereby reducing $p$), they can invest in capabilities to detect misuse or impose greater penalties on identified malicious actors (thereby increasing $P\lambda$),\footnote{``Increasing penalties'' here can mean anything that imposes additional friction upon propagandists once identified. For instance, requiring a CAPTCHA in addition to an email address for users signing up for model access imposes additional costs, though not very large ones.} or they can embed watermarks into model outputs or pursue other strategies that increase the potential for detection of synthetic content online.\footnote{The model presented here does not readily include a way for analyzing this strategy. Watermarks reduce the value of LLM outputs, but they do not make it more costly to generate the outputs, meaning that a strict cost comparison does not capture the relevant differences.} The model and analysis presented here indicates that all three strategies can be valuable, though in different ways.

Model alignment efforts that reduce $p$ and monitoring controls that increase $P\lambda$ are primarily useful when rival open source models do not exist, or if propagandists are particularly ``sticky'' and unlikely to switch to such rival models. Model alignment efforts can also be pursued by groups who develop and release open source models, though to date, this is less common than among businesses that seek to monetize their models. However, monitoring controls are not in principle applicable to open source models (at least not in the form analyzed here, though see footnote \ref{opensourcemonitoring}).\footnote{Note that even if these interventions do not carry major benefits from a security perspective, they may still be valuable from a safety perspective. In addition, if propagandists are satisficers, it is possible that a failed attempt to make use of an API-accessible model (whether due to the model's refusal to produce the desired outputs or a quick detection) may dissuade them from seriously pursuing the use of LLMs by other means as well.}

The development of watermarks for LLMs is an active area of research. Existing proposals for technical methods of watermarking LLMs, however, are ``shallow'' in the sense that they are added on top of a pretrained LLM and can easily be removed by a user or eliminated via fine-tuning. \cite{watermark} However, a growing number of researchers are exploring the feasibility of ``deep'' watermarks or other methods that persist and allow for attribution even after fine-tuning. \cite{deepwatermark, attribution, competition} Whether or not such deep watermarks will prove to be feasible is an open question—but if they are, and if propagandists looking to use LLMs for content generation primarily rely on fine-tuned versions of open source models, then embedding such watermarks into open source models may be a valuable intervention. 

It is important to emphasize, however, that while the model presented here can shed some light on the expected value of some types of technical interventions, no combination of strictly technical safeguards is likely to fully address the issues posed by LLM-enabled influence operations. To more comprehensively address these risks, it will be necessary to combine technical, policy, and legal interventions. For more discussion about the variety of technical, policy, and legal tools available to various stakeholders, see (\textit{inter alia}) \cite{workshop, sedova1, sedova2, shelvane, solaiman, releasestrats, deepmind}. 

\section*{Acknowledgements and Supplemental Materials}

For their comments, discussion, careful analysis, and general supportiveness for this work, I would like to thank Renée DiResta, Irene Solaiman, Katherine Quinn, Girish Sastry, Josh Goldstein, Andrew Lohn, Mia Hoffmann, and John Bansemer. All errors remain my own. 

A GitHub repository for this work, which contains code to produce all associated figures and results via Monte Carlo estimation, is available at \href{https://github.com/georgetown-cset/disinfo-costs}{https://github.com/georgetown-cset/disinfo-costs}. A blog-style summary of the work for policymakers can be found at \href{https://cset.georgetown.edu/article/how-much-money-could-large-language-models-save-propagandists/}{https://cset.georgetown.edu/article/how-much-money-could-large-language-models-save-propagandists/}.

\newpage 

\appendix
\section{Supplemental Equations}
\label{sec:supplemental}

The model presented in this paper requires only algebraic manipulation, primarily of equation \ref{eq:totalcosts}, in order to calculate any of the final variables of interest discussed in Section \ref{sec:robustness}. However, some of this algebraic manipulation can be tedious, so I reproduce here some useful solutions for various parameters of interest. 

First, let $\hat{p}_{1H}$ represent the threshold performance value at which the use of the private model is preferred to reliance on a manual campaign (assuming the detection rate $\lambda$ is already fixed). Then this value is given by:

\begin{equation}
  \hat{p}_{1H} = \frac{1}{\alpha} + \frac{L}{w} \left( IC + P \lambda \right)
\end{equation}

Let $\hat{p}_{1AI}$ represent the threshold performance value at which the use of the private model is preferred to an alternative open source option. This value is given by:

\begin{equation}
  \hat{p}_{1AI} = p_2 + \frac{p_2 P \lambda - \frac{FC}{n}}{\frac{w}{\alpha L} + IC} 
\end{equation}

Note that, if $p_1 = p_2$, the propagandist is only indifferent between the API-accessible and open source models if $p_2 P \lambda = \frac{FC}{n}$. In other words, the API-accessible model can perform worse than the open source model and face meaningful monitoring risks, and yet still be preferred if $\frac{FC}{n}$ is sufficiently large.  

If, alternatively, values of $p_1$ are already known, we may instead want to calculate the minimum detection capability at which a propagandist is deterred from using the API-acccessible model. If the propagandist's fallback to the use of the API-accessible model is a manual campaign, then the minimum deterrant detection capability $\hat{\lambda}_H$ is given by:  

\begin{equation}
  \hat{\lambda}_{H} = \frac{1}{P} \left( \frac{w p_1}{L} - \frac{w}{\alpha L} - IC \right)
\end{equation}

Alternatively, if the public model is sufficiently well-performing that the propagandist will use it instead as a fallback, then the minimum deterrant detection capability $\hat{\lambda}_{AI}$ is instead given by:

\begin{equation}
  \label{eq:lambdahatAI}
  \hat{\lambda}_{AI} = \frac{1}{P} \left( \frac{w}{\alpha L} + IC \right) \left( \frac{p_1 - p_2}{p_2} \right) + \frac{p_1 FC}{n P}
\end{equation}

Note that in the two preceding equations, a detection penalty of \$0 causes $\hat{\lambda}$ to be undefined, because no detection capability could possibly deter malicious use if the detection does not itself impose some form of penalty. In addition, note that the equation \ref{eq:lambdahatAI} is the general case of equation \ref{eq:public2}, where fixed costs associated with running an open source model are no longer assumed to be \$0.

Finally, let equation \ref{eq:totalcosts} represent the three choices for operation facing the propagandist, but let it also be possible for the propagandist to spend $FC$ (or $\Delta FC$ more than was spent for the use of the current open source model option) to create or fine-tune a new model with target capability $\hat{p}$. Then, for each of the three campaign styles, we can set $\frac{n}{\hat{p}} \left( \frac{w}{\alpha L} + IC \right) + FC$ equal to the corresponding portion of equation \ref{eq:public2} and solve for $n$ to calculate the minimum campaign size at which expending $FC$ becomes cost effective relative to an existing choice of model (including manual authorship as a ``choice'' of model). The overall minimum viable scale for a model where $(FC, \hat{p}) \in FC$ is then the maximum solution of $n$ across all alternative model choices (because the minimum viable scale is the scale at which a model becomes cost effective relative to the next-most-cost-effective option). This is given by:

\begin{equation}
  \hat{n} = \max \left(
    \begin{aligned}
      \text{Manual} &= \frac{\hat{p} FC}{\frac{\hat{p} w}{L} - \left( \frac{w}{\alpha L} + IC \right)} \\
      \text{API-Accessible LLM} &= \frac{(\hat{p} \cdot p_1) FC}{\hat{p} P \lambda + (\hat{p} - p_1) \left( \frac{w}{\alpha L} + IC \right)} \\
      \text{Open Source LLM} &= \frac{(\hat{p} \cdot p_2) \Delta FC}{(\hat{p} - p_2) \left( \frac{w}{\alpha L} + IC \right)}
    \end{aligned}
  \right)
\end{equation}

\end{document}